\documentclass[12pt]{article}
\usepackage{amsmath}
\usepackage{graphicx,psfrag,epsf}
\usepackage{enumerate}
\usepackage{natbib}
\usepackage{url} 
\usepackage{enumitem} 
\usepackage{multirow}
\usepackage[font={stretch=0.7}]{caption} 
\usepackage{bbm}
\usepackage{float}
\usepackage{booktabs,makecell}
\usepackage{amsthm,amssymb}
\usepackage{amsfonts}
\usepackage{dsfont}
\usepackage{amsmath}
\usepackage{amssymb}
\usepackage{hyperref}
\usepackage{multirow}
\usepackage{booktabs}
\usepackage{mathrsfs} 
\usepackage{upgreek}
\hypersetup{
     colorlinks   = true,
  linkcolor=blue,
            urlcolor=blue,
            citecolor=blue
}
\usepackage{xr}
\makeatletter
\newcommand*{\addFileDependency}[1]{
  \typeout{(#1)}
  \@addtofilelist{#1}
  \IfFileExists{#1}{}{\typeout{No file #1.}}
}
\makeatother

\newcommand*{\myexternaldocument}[1]{%
    \externaldocument{#1}%
    \addFileDependency{#1.tex}%
    \addFileDependency{#1.aux}%
}

\myexternaldocument{Supplementary}
\usepackage[ruled,vlined,onelanguage]{algorithm2e}

\newcommand{\blind}{0}
  
\begin{document}

\def\spacingset#1{\renewcommand{\baselinestretch}%
{#1}\small\normalsize} \spacingset{1}

\def\bmZ{\boldsymbol{Z}}
\def\bmY{\boldsymbol{Y}}
\def\bmW{\boldsymbol{W}}
\def\bmM{\boldsymbol{M}}
\def\bmE{\boldsymbol{\mathcal{E}}}
\def\bmlambda{\boldsymbol{\lambda}}
\def\bmrho{\boldsymbol{\rho}}
\def\Cov{{\rm Cov}}
\def\rmvec{{\rm vec}}
\def\bmSigma{\boldsymbol{\Sigma}}
\def\mbR{\mathbb{R}}
\def\mbW{\mathbb{W}}
\def\bmI{\boldsymbol{I}}
\def\bmy{\boldsymbol{y}}
\def\bmw{\boldsymbol{w}}
\def\bmvarepsilon{\boldsymbol{\varepsilon}}
\def\bmxi{\boldsymbol{\xi}}
\def\hbmbeta{\boldsymbol{\hat\beta}}
\def\tbmbeta{\boldsymbol{\tilde\beta}}
\def\bbmbeta{\boldsymbol{\bar\beta}}
\def\mL{\mathscr{L}}
\def\tmL{\widetilde {\mL}}
\def\bmeta{\boldsymbol{\eta}}
\def\lbk{\left\{}
\def\rbk{\right\}}
\def\lak{\left|}
\def\rak{\right|}
\def\lmk{\left[}
\def\rmk{\right]}
\def\lsk{\left(}
\def\rsk{\right)}
\def\bmx{\boldsymbol{x}}
\def\bmz{\boldsymbol{z}}
\def\bmv{\boldsymbol{v}}
\def\bmmu{\boldsymbol{\mu}}
\def\bmzeta{\boldsymbol{\zeta}}
\def\bmsigma{\boldsymbol{\sigma}}
\def\bmbeta{\boldsymbol{\beta}}
\def\var{{\rm var}}
\def\bml{\boldsymbol{1}}
\def\bmo{\boldsymbol{0}}
\def\diag{{\rm diag}}
\def\bmrho{\boldsymbol{\rho}}
\def\E{{\rm E}}
\def\bmgamma{\boldsymbol{\gamma}}
\def\bmalpha{\boldsymbol{\alpha}}
\def\bmvtheta{\boldsymbol{\vartheta}}
\def\laak{\left\|}
\def\raak{\right\|}
\def\OLS{{\rm OLS}}
\def\GLS{{\rm GLS}}
\def\FGLS{{\rm FGLS}}
\def\LS{{\rm LS}}
\def\FLS{{\rm FLS}}
\def\bmtheta{\boldsymbol{{\mathcal{B}}}}
\def\tr{{\rm tr}}
\def\mbY{\mathbb{Y}}
\def\mbX{\mathbb{X}}
\def\rmE{{\rm E}}
\def\rmP{{\rm P}}
\def\Cov{{\rm Cov}}
\def\Var{{\rm Var}}
\def\mS{{\mathcal{S}}}
\def\mU{{\mathcal{U}}}
\def\mN{{\mathcal{N}}}
\def\mP{{\mathcal{P}}}
\def\mV{{\mathcal{V}}}
\def\sgn{\operatorname{sgn}}
\def\bu{\boldsymbol u}

\def\be{\begin{equation}}
\def\ee{\end{equation}} 
\def\ben{\begin{equation*}}
\def\een{\end{equation*}}
\def\bea{\begin{eqnarray}}
\def\eea{\end{eqnarray}}
\def\bda{\begin{eqnarray*}}
\def\eda{\end{eqnarray*}}
\numberwithin{equation}{section}
\def\m{\color{magenta}}
\def\b{\color{blue}}
\def\here{{\color{magenta}Tao is up to here.}}
\def\XLhere{{\color{magenta}Xian is up to here.}}

\newtheorem{theorem}{Theorem}
\newtheorem{proposition}{Proposition}
\newtheorem{corollary}{Corollary}
\newtheorem{assumption}{Assumption}
\newtheorem{lemma}{Lemma}
\newtheorem{remark}{Remark}


\if0\blind
{\spacingset{1.5} 
  \title{\bf Subbagging  Variable Selection for Big Data}
  \date{}
  \author{Xian Li, Xuan Liang and Tao Zou 
 \\
    \textit{The Australian National University
    }
    } 
  
  \maketitle

} \fi

\bigskip
\begin{abstract}
{

This article introduces a subbagging (subsample aggregating) approach for variable selection in regression within the context of big data. The proposed subbagging approach not only ensures that variable selection is  scalable given the constraints of available computational resources, but also preserves the statistical efficiency of the resulting estimator. In particular, we propose a subbagging loss function that aggregates the least-squares approximations of the loss function for each subsample. Subsequently, we penalize the subbagging loss function via an adaptive LASSO-type regularizer, and obtain a regularized estimator to achieve variable selection. We then demonstrate that the regularized estimator exhibits $\sqrt{N}$-consistency and possesses the oracle properties, where $N$ represents the size of the full sample in the big data.   In addition, we propose a subbagging Bayesian information criterion to select the regularization parameter, ensuring that the regularized estimator achieves selection consistency.
Simulation experiments are conducted to demonstrate the numerical performance. A U.S. census dataset is analyzed to illustrate the usefulness and computational scalability of the subbagging variable selection method.

}
\spacingset{1.2}

\end{abstract}

\noindent%
{{\it Keywords:} Bayesian Information Criterion, Computational Scalability, Selection Consistency,  Subsample Aggregation, Regularization.
} 
\vfill

\newpage
\spacingset{1.5} 

\section{Introduction}
\label{sec:intro}

With the rapid advancement of technology, information and data collection have become easier than ever, propelling us into the era of big data.
 Massive data sizes have presented both opportunities and challenges for modern society and data scientists. On the one hand, enormous data sizes enrich our information landscape, enabling the discovery of subtle population patterns that cannot be achieved by analyzing small datasets. On the other hand, the sheer volume and complexity of big data introduce significant computational challenges (\citealp{fan2014challenges}). For instance, practitioners may not be able to 
 directly analyze massive datasets due to the constraints of available computational resources (\citealp{meng2018statistical}).

To tackle the challenges, a widely used approach is to select a subsample to represent the full massive dataset, a method commonly known as subsampling. Relevant studies in this area include \cite{drineas2006sampling}, \cite{drineas2011faster}, \cite{fithian2014local}, \cite{ma2015statistical} and \cite{wang2018optimal}. 
Although subsampling is computationally scalable when faced with limited computational resources, reducing the dataset's size inevitably leads to a loss of statistical efficiency and can result in misleading outcomes due to potential subsample biases.

In order to address the above issues, the literature has explored subsample aggregating, also known as ``subbagging'' as introduced by \cite{buhlmann2002analyzing}, which involves aggregating statistical analysis results from multiple subsamples.    
 Other related literature may employ different terminologies, such as ``combining sketches" introduced in \cite{lee2020econometric}.  
During the past few years, subbagging has attracted considerable attention, with focuses mostly on prediction and estimation. For instance, \cite{andonova2002simple} developed a subbagging algorithm to improve the predictive performance of classification, while \cite{zaman2009effect}  investigated how the subsample size within the procedure of subbagging impacts on the prediction accuracy. Both of them demonstrated that subbagging outperforms traditional bagging in various scenarios. \cite{mentch2016quantifying} integrated subbagging into random forests predictions and examined their theoretical properties. \cite{lee2020econometric} explored the impact of subbagging on prediction and estimation within a linear regression setup. Furthermore, \cite{lee2020sketching} investigated the properties of  subbagging  for least-squares estimation when subsamples are generated by a sampling scheme called ``countsketch''. \cite{zou2021subbagging} introduced the consistency and asymptotic normality of the subbagging $Z$-estimator, targeting a general estimation problem that involves solving estimating equations.

In addition to prediction and estimation, variable selection is also an important part of statistical analysis. For instance, effective variable selection techniques can facilitate the discovery of important variables. Over the past few decades, numerous variable selection methods have been actively investigated (see, e.g., \citealp{tibshirani1996regression}, \citealp{fan2001variable}, \citealp{zou2006adaptive} and \citealp{zhang2010nearly}).
However, to the best of our knowledge, there is limited research on how to properly conduct variable selection within the realm of subbagging. One possible reason is that directly incorporating the subbagging approach into variable selection requires aggregating the sets of selected variables obtained from multiple subsamples. These sets can differ significantly due to the variation among different subsamples, thereby complicating the aggregation of subsample variable selection results. 
Some related literature has explored aggregating the sets of selected variables using methods such as intersection or union (\citealp{beinrucker2016extensions}), a forward aggregation approach (\citealp{bolon2015distributed}), and majority voting of variables (\citealp{ibarguren2022pctbagging} and \citealp{capanu2023subsampling}). However, there are several limitations in these approaches. 
First, the intersection approach can under-select variables,  resulting in an empty set in the worst-case scenario, while the union approach can over-select variables. 
Second, the forward aggregation in subbagging is similar to the forward selection approach which only allows adding but not removing variables. 
Lastly, the majority voting approach is based on the frequency with which a variable is selected across multiple subsample variable selection results, but the cutoff frequency used to define the majority is arbitrary.  From the theoretical perspective, only \cite{capanu2023subsampling} established screening consistency (no false exclusion) for their approach, but not variable selection consistency, while other literature offers no theoretical guarantees.

In this article, we propose a novel subbagging variable selection approach that differs from existing methods and is applicable to a wide range of regression problems. Specifically, our approach is different from the aforementioned literature in that we do not directly aggregate the sets of selected variables from multiple subsamples. Instead, we propose to aggregate some statistics, other than the variable selection results, from each subsample, the aggregation of which is then used to achieve simultaneous variable selection and parameter estimation. In particular, our approach is shown to exhibit $\sqrt{N}$-consistency and the oracle properties (see, e.g., \citealp{fan2001variable},  \citealp{zou2006adaptive} and \citealp{wang2007unified}) for estimation, and it possesses selection consistency for variable selection, where $N$ represents the size of the full sample in the big data. 
It is worth noting that our approach requires only a few statistics obtained by fitting a single model for each subsample. This is different from \cite{capanu2023subsampling}, where they require fitting a sequence of nested models. Consequently, our approach is computationally less expensive. Finally, since our approach is based on subbagging, which originates from subsampling, it naturally inherits the advantage of being computationally scalable given the constraints of available computational resources.

In the following, we provide a sketch of our proposed subbagging variable selection approach. For a general regression problem, let $\bmbeta\in\mathbb{R}^p$ be the $p$-dimensional vector of regression coefficients, which are associated to $p$ covariates. Our method consists of two steps given below.

Step 1: We randomly draw $m_N$ subsamples independently from the full sample. Particularly, each subsample $s$ consists of $k_N$ distinct observations from the full dataset and leads to an estimator ${\tbmbeta}_{k_N,s}$ by minimizing the loss function for each subsample. Subsequently, we use ${\tbmbeta}_{k_N,s}$ to obtain the least-squares approximation \citep{wang2007unified} of each subsample loss function. It is worth noting that the least-squares approximation is a quadratic function that is only determined by several statistics constructed from  ${\tbmbeta}_{k_N,s}$ and the subsample itself.

 Step 2: We aggregate the least-squares approximations of $m_N$ subsample loss functions,  and obtain an overall loss function referred to as the subbagging loss function. We next penalize this function via an adaptive LASSO-type regularizer, and obtain a regularized estimator which achieves both  variable selection and parameter estimation.

It is worth noting that Step 1 can be executed in parallel, which can reduce the computation time for this step. In Step 2,  the aggregation of  the least-squares approximations can be easily realized by  averaging the several statistics obtained from Step 1,  rather than the original subsamples, making this step computationally efficient; see the details in Section \ref{sec:ensemble}. This also avoids aggregating the sets of selected variables considered in the existing literature. 
To ensure the selection consistency achievable by the regularized estimator, 
we propose a subbagging Bayesian information criterion (SBIC) to select the regularization parameter. Note that the subbagging loss function cannot be used directly to construct the SBIC; it needs to be scaled by $k_N$. Detailed explanations are provided in Section \ref{sec:SBIC}. 
Extensive simulation studies and a case study regarding U.S. census data are conducted to demonstrate the computational scalability and the usefulness of the regularized estimator and SBIC.

The rest of the article is organized as follows. Section \ref{sec:ensemble} introduces the framework of subbagging and elaborates on the method of subbagging variable selection. Section \ref{sec:theo} presents the theoretical properties of our proposed method. Simulation studies are provided in Section \ref{sec:simu}, and a real data analysis to demonstrate the usefulness of our method is given in Section \ref{sec:real}. Section \ref{sec:co} concludes this article with a discussion. Technical conditions  are provided in Appendix.  Proofs of the theorems, useful lemmas and additional numerical results are relegated to the supplementary material.

\section{Subbagging Variable Selection}
\label{sec:ensemble}


Recall that in Introduction, we denote $N$ as the size of the full sample in the big data. Subsequently, we let the full sample be $\{\bmz_i=(y_i,\bmx_i^\top)^\top:i=1,\cdots,N\}$,  where $y_i\in \mathbb{R}$ is the response variable and $\bmx_i^\top \in \mathbb{R}^p$ is the covariate vector. In addition, we consider that $\bmz_i$, $i=1,\cdots, N$, are independent and identically distributed. 

In this paper, we focus on the variable selection under the linear model or generalized linear model framework. Specifically, the regression coefficient vector $\bmbeta=(\beta_1,\cdots,\beta_p)^\top$, which is associated to the $p$ variables $x_{i1},\cdots,x_{ip}$ in the covariate vector $\bmx_i=(x_{i1},\cdots,x_{ip})^\top$, can be estimated via
 minimizing the full sample loss function  $\mL_{N}(\bmbeta)= N^{-1} \sum_{i=1}^N
\mL(\bmbeta;\bmz_i)$, where $\mL(\bmbeta;\bmz_i)$ is  $(y_i-\bmx_i^\top\bmbeta)^2$ for the linear model, and is the negative log-likelihood (or quasi-log-likelihood) function of the $i$-th observation $\bmz_i = (y_i, \bmx_i^\top)^\top$ for the generalized linear model. Subsequently, we obtain the full sample estimator  $\tbmbeta_N=\arg\min_{\bmbeta} \mL_{N}(\bmbeta)$. Various variable selection methods can also be applied based on the full sample and the loss function $\mL_{N}(\bmbeta)$. For example, \cite{zou2006adaptive} proposes the  adaptive LASSO approach which is to minimize the following regularized loss function:
\be\mL_N(\bmbeta)+\lambda\sum_{j=1}^pw_j|\beta_j|,\label{global:al}\ee
where $\lambda$ is the regularization parameter and $w_j$ is the adaptive weight applied to each regression coefficient $\beta_j$. \cite{zou2006adaptive} has further shown that for the  linear model,  the regularized estimator that minimizes (\ref{global:al}) can achieve variable selection consistency under some mild conditions.

As discussed in Introduction, obtaining $\tbmbeta_N$ and performing variable selection based on the full sample can be computationally challenging when $N$ is exceedingly large. 
To this end, we consider the subbagging approach in the literature.  Specifically, we draw $m_N$ subsamples independently from the full sample $\{1, \dots, N\}$, and each subsample $s \subset \{1, \dots, N\}$ consists of $k_N$ observation indices $s_1, \dots, s_{k_N}$ sampled without replacement. Accordingly,  for  each subsample $s=\{s_1,\cdots,s_{k_N}\}$, its size 
is $k_N$, its associated subsample loss function is 
\be 
\label{eq:subloss} 
\mL_{k_N,s}(\bmbeta)=\frac{1}{k_N} \sum_{i\in s} \mL(\bmbeta;\bmz_i)\nonumber,
\ee  and the  subsample estimator of $\bmbeta$ can be obtained by
\be 
\label{eq:subest} 
\tbmbeta_{k_N,s}=\underset{\bmbeta}{\arg\min}\ \mL_{k_N,s}(\bmbeta).\nonumber
\ee
Since the subsample size $k_N$ can be chosen to be much smaller than the full sample size $N$, obtaining each $\tbmbeta_{k_N,s}$ becomes computationally more feasible. 
For example, the full dataset in our real data analysis of Section \ref{sec:real} consists of $N=15,965,200$ observations and occupies approximately 8 gigabytes (GB) on the hard drive. In contrast, a subsample of size $k_N=252,569$ occupies only around 128 megabytes (MB), which makes it easily loadable into memory for analysis, including calculating the subsample estimator $\tbmbeta_{k_N,s}$.

Next, we propose our subbagging variable selection approach. 
In the subbagging framework, we randomly draw $m_N$ subsamples, and hence we can collect them into a set $\mS = \{s^{(1)}, \cdots, s^{(m_N)}\}$. 
Based on the idea of subbagging (subsample aggregating), we can aggregate the $m_N$ subsample loss functions $\mL_{k_N,s}(\bmbeta)$ to obtain an approximation of the full sample loss function $\mL_N(\bmbeta)$, which is $m_N^{-1} \sum_{s \in \mS} \mL_{k_N,s}(\bmbeta)$. Hence, if we replace $\mL_N(\bmbeta)$ in (\ref{global:al}) with $m_N^{-1} \sum_{s \in \mS} \mL_{k_N,s}(\bmbeta)$, the resulting regularized estimator should also achieve variable selection consistency. However, for any given $\bmbeta$, the computation of $m_N^{-1} \sum_{s \in \mS} \mL_{k_N,s}(\bmbeta)$ needs the knowledge of all $m_N$ subsamples $\mS = \{s^{(1)}, \cdots, s^{(m_N)}\} = \big\{\{s_1^{(1)}, \cdots, s_{k_N}^{(1)}\}, \cdots, \{s_1^{(m_N)}, \cdots, s_{k_N}^{(m_N)}\}\big\}$, which requires saving all $m_N k_N$ observations associated to $m_N^{-1} \sum_{s \in \mS} \mL_{k_N,s}(\bmbeta)$ into the memory. This is, again, not computationally feasible, when $m_N k_N$ is large.

To tackle this problem, we adapt the least-squares approximation (see \citealp{wang2007unified} and \citealp{zhu2021least}) to obtain a further approximation of $m_N^{-1} \sum_{s \in \mS} \mL_{k_N,s}(\bmbeta)$.
 Specifically, we consider the least-squares approximation for each subsample loss function $\mL_{k_N,s}(\bmbeta)$. Let the first and second-order derivatives of $\mL(\bmbeta;\bmz_i)$ be $\nabla\mL(\bmbeta;\bmz_i)=\partial \mL(\bmbeta;\bmz_i)/\partial \bmbeta$ and $\nabla^{2}\mL(\bmbeta;\bmz_i)=\partial^2 \mL(\bmbeta;\bmz_i)/(\partial \bmbeta\partial \bmbeta^\top)$, respectively.
Accordingly, $\nabla\mL_{k_N,s}$ $(\bmbeta)= k_N^{-1} \sum_{i\in s}
\nabla\mL(\bmbeta;\bmz_i)$ and $\nabla^{2}\mL_{k_N,s}(\bmbeta)= k_N^{-1} \sum_{i\in s}
\nabla^{2}\mL(\bmbeta;\bmz_i)$. Suppose $\bmbeta$ is near $\tbmbeta_{k_N,s}$ and consider 
 Taylor's  expansion of $\mL_{k_N,s}(\bmbeta)$ at  $\tbmbeta_{k_N,s}$, which leads to 
\bea
\mL_{k_N,s}(\bmbeta)&=&\mL_{k_N,s}\lsk\tbmbeta_{k_N,s}\rsk+\frac{1}{2} \lsk\bmbeta-\tbmbeta_{k_N,s}\rsk^{\top} \nabla^{2}\mL_{k_N,s}\lsk\tbmbeta_{k_N,s} \rsk\lsk\bmbeta-\tbmbeta_{k_N,s}\rsk\nonumber\\&&+o_{\rmP}\lsk \laak \bmbeta-\tbmbeta_{k_N,s}\raak_2^2 \rsk\label{eq:subsample_loss_TS},
\eea
where $\laak\cdot\raak_2$ is the vector 2-norm. This result is obtained due to that $\tbmbeta_{k_N,s}$ solves $\nabla\mL_{k_N,s}(\tbmbeta_{k_N,s})$ $=\boldsymbol 0_p$, where 
 $\boldsymbol 0_p$ is the $p$-dimensional  vector of zeros. 
Ignoring the smaller order terms, we then approximate 
$m_N^{-1} \sum_{s \in \mS} \mL_{k_N,s}(\bmbeta)$ by
\bda
&&\frac{1}{m_N}\sum_{s\in\mS}\lbk \mL_{k_N,s}\lsk\tbmbeta_{k_N,s}\rsk
+\frac{1}{2} \lsk\bmbeta-\tbmbeta_{k_N,s}\rsk^{\top} \nabla^{2}\mL_{k_N,s}\lsk\tbmbeta_{k_N,s} \rsk\lsk\bmbeta-\tbmbeta_{k_N,s}\rsk\rbk\\
&=&\frac{1}{2m_N}\sum_{s\in\mS}\lsk\bmbeta-\tbmbeta_{k_N,s}\rsk^{\top} \nabla^{2}\mL_{k_N,s}\lsk\tbmbeta_{k_N,s} \rsk\lsk\bmbeta-\tbmbeta_{k_N,s}\rsk+C_\mL\\
&=&\frac{1}{2}\tmL_{k_N,m_N}(\bmbeta)+C_\mL,
\eda
where $C_\mL={m_N}^{-1}\sum_{s\in\mS}\mL_{k_N,s}(\tbmbeta_{k_N,s})$ is a constant not depending on $\bmbeta$, and 
\be\label{lsa:def}
\tmL_{k_N,m_N}(\bmbeta)=\frac{1}{m_N}\sum_{s\in\mS}\lsk\bmbeta-\tbmbeta_{k_N,s}\rsk^{\top} \nabla^{2}\mL_{k_N,s}\lsk\tbmbeta_{k_N,s} \rsk\lsk\bmbeta-\tbmbeta_{k_N,s}\rsk
\ee
is referred to as the subbagging loss function in this paper.

In contrast to $m_N^{-1} \sum_{s \in \mS} \mL_{k_N,s}(\bmbeta)$, the computation of the subbagging loss function $\tmL_{k_N,m_N}(\bmbeta)$ only needs the knowledge of the statistics $\{\big(\tbmbeta_{k_N,s}, \nabla^{2}\mL_{k_N,s}(\tbmbeta_{k_N,s} )\big):s\in\mS\}$, which requires saving  $m_N$ observations in $\{\big(\tbmbeta_{k_N,s}, \nabla^{2}\mL_{k_N,s}(\tbmbeta_{k_N,s} )\big):s\in\mS\}$ into the memory. Hence, the computation of $\tmL_{k_N,m_N}(\bmbeta)$ is much more feasible than that of $m_N^{-1} \sum_{s \in \mS} \mL_{k_N,s}(\bmbeta)$. In fact, to compute $\tmL_{k_N,m_N}(\bmbeta)$ in the subbagging procedure, we only need to retain the statistics $\{\big(\tbmbeta_{k_N,s}, \nabla^{2}\mL_{k_N,s}(\tbmbeta_{k_N,s} )\big) : s \in \mS\}$, allowing us to fully discard the original subsamples and significantly reduce memory usage. In addition, since the $m_N$ subsamples in $\mS$ are drawn independently, computing the statistics $\{\big(\tbmbeta_{k_N,s}, \nabla^{2}\mL_{k_N,s}(\tbmbeta_{k_N,s} )\big) : s \in \mS\}$ can be performed in parallel, which further reduces computation time.




Due to the computational advantage of the subbagging loss function $\tmL_{k_N,m_N}(\bmbeta)$, we replace $\mL_N(\bmbeta)$ in (\ref{global:al})  with $\tmL_{k_N,m_N}(\bmbeta)$ , and obtain the  regularized subbagging estimator
\be\label{loss:reg}
\hbmbeta_{k_N,m_N}=\underset{\bmbeta}{\arg\min} \lbk \tmL_{k_N,m_N}(\bmbeta)+\lambda\sum_{j=1}^pw_j|\beta_j|\rbk.
\ee
In practice, 
the adaptive weight can be chosen to be $w_j=1/|\tilde\beta_{k_N,m_N,j}|^\gamma$ for some $\gamma>0$, where $\tilde\beta_{k_N,m_N,j}$ is the $j$-th element of the subbagging estimator  $\tbmbeta_{k_N,m_N}=m_N^{-1}\sum_{s\in\mS}\tbmbeta_{k_N,s}$ (see \citealp{zou2021subbagging}), for $j=1,\cdots, p$. In our numerical studies of Sections \ref{sec:simu} -- \ref{sec:real}, we set $\gamma=1$ as suggested by  \cite{zou2006adaptive}. 

Let $\bmbeta_0=(\beta_{0,1},\cdots,\beta_{0,p})^\top$ be the true parameter vector of $\bmbeta$. Under Conditions (C1) -- (C5) in Appendix, Lemma \ref{ps:3} of the supplementary material shows that $\tbmbeta_{k_N,m_N}$ is  $\sqrt{N}$-consistent to $\bmbeta_0$. Accordingly, if $\beta_{0,j}=0$ for some $j\in\{1,\cdots,p\}$,  then $w_j=1/|\tilde\beta_{k_N,m_N,j}|^\gamma$ grows to infinity in probability as $N\to\infty$; otherwise, $w_j=1/|\tilde\beta_{k_N,m_N,j}|^\gamma$ converges to a finite positive constant in probability. As a consequence, the zero regression coefficient $\beta_{0,j}$ leads to a large weight $w_j=1/|\tilde\beta_{k_N,m_N,j}|^\gamma$ in the regularizer, ensuring that the estimation of $\hat\beta_{k_N,m_N,j}$ (the $j$-th element of $\hbmbeta_{k_N,m_N}$) in (\ref{loss:reg}) shrinks rapidly to zero. This then achieves variable selection because $\beta_{0,j}=0$ corresponds to the irrelevant variable $x_{ij}$ in the covariate vector $\bmx_{i}$. In the following section, we will demonstrate that the  regularized subbagging estimator $\hbmbeta_{k_N,m_N}$ in (\ref{loss:reg}) not only possesses variable selection consistency but also exhibits $\sqrt{N}$-consistency and oracle properties, given proper rate conditions for the regularization parameter $\lambda$.



\section{Theoretical Results}\label{sec:theo}

\subsection{Asymptotic Theory for Regularized Subbagging Estimator}\label{subsec:AT}

Before presenting our theoretical results, we introduce some notation.  Without loss of generality,  we assume that the first $p_0$ elements in the true parameter vector $\bmbeta_0$ are nonzero, while the remaining  $p-p_0$ elements are zeros, where $0<p_0 < p$. Correspondingly, let $\mathcal{M}_T = \{1, \cdots, p_0\}$ represent the true model, which is associated to the relevant variables $x_{i1},\cdots,x_{ip_0}$ in the covariate vector $\bmx_i$, and let $\mathcal{M}_F=\{1,\cdots,p\}$ be the full model. Moreover,  for any $p$-dimensional vector $\bmv=(v_1,\cdots, v_p)^\top$ and any set $\mathcal{M}=\{j_1,\cdots, j_k\}\subset \mathcal{M}_F$, denote $\bmv^{(\mathcal{M})}=(v_{j_1},\cdots, v_{j_k})^\top$, a vector that extracts the elements of $\bmv$ based on the indices in $\mathcal{M}$. In addition, we define $\bmv^{(-\mathcal{M})}=\bmv^{\mathcal{M}_F\backslash\mathcal{M}}\in\mathbb{R}^{p-k}$. Next, for any generic matrix $G=(g_{\ell_1 \ell_2})_{\ell_1,\ell_2\in\mathcal{M}_F}\in \mathbb{R}^{p\times p}$ with its $(\ell_1,\ell_2)$-th element being $g_{\ell_1\ell_2}$, define $G^{(\mathcal{M})} = (g_{\ell_1 \ell_2})_{\ell_1,\ell_2\in \mathcal{M}} \in \mathbb{R}^{k\times k}$ as a submatrix of $G$ obtained by extracting its $j_1$-th, $\cdots$, $j_k$-th rows and $j_1$-th, $\cdots$, $j_k$-th columns. Furthermore, 
denote $\Sigma_{\bmbeta_0}=\Var\{\nabla\mL(\bmbeta_0;\bmz_i)\} \in\mathbb{R}^{p\times p}$, which is the variance-covariance matrix of  $\nabla\mL(\bmbeta_0;\bmz_i)$, 
and $V_{\bmbeta_0}=\rmE\{\nabla^{2}\mL(\bmbeta_0;\bmz_i)\}\in\mathbb{R}^{p\times p}$. Finally, let $a_w = \max\{\lambda w_j :1\leq j \leq p_0\}$ and $b_w = \min\{\lambda w_j :p_0<j \leq p\}$.  When we obtain the  regularized subbagging estimator $\hbmbeta_{k_N,m_N}$ in (\ref{loss:reg}), $a_w$ controls the largest
penalty on the first $p_0$ elements of $\bmbeta$, while $b_w$ controls the smallest penalty on the last $p-p_0$ elements of $\bmbeta$.   
 
Based on the  notation, we then establish the asymptotic properties of the  regularized subbagging estimator $\hbmbeta_{k_N,m_N}$ given below.

\begin{theorem}\label{tm:1}Under Conditions  {(C1) -- (C4)}   in Appendix, assume $k_N\to\infty$, $k_N/N\to 0$, $k_N/\sqrt{N}\to\infty$ and $k_Nm_N/N\to\alpha\in(0,\infty]$  as $N\to \infty$.  Then, we obtain the following results.

(1) ($\sqrt{N}$-Consistency) If $\sqrt{N}a_w\stackrel{\rmP}{\longrightarrow} 0$, then \[ \hbmbeta_{k_N,m_N} -\boldsymbol{ \beta}_{0}=O_{\rmP}\lsk N^{-1/2}\rsk .\]

(2) (Selection Consistency) If $\sqrt{N}a_w\stackrel{\rmP}{\longrightarrow} 0$ and $\sqrt{N}b_w\stackrel{\rmP}{\longrightarrow} \infty$,  then 
\[\rmP\lsk\hbmbeta_{k_N,m_N}^{(-\mathcal{M}_T)}=\boldsymbol {0}_{p-p_0}\rsk\to1.\]

(3) (Asymptotic Normality) If $\sqrt{N}a_w\stackrel{\rmP}{\longrightarrow} 0$, $\sqrt{N}b_w\stackrel{\rmP}{\longrightarrow} \infty$ and Condition (C5) in Appendix  hold,  then\[
\sqrt{N}\lsk\hbmbeta_{k_N,m_N}^{(\mathcal{M}_T)}-\bmbeta_0^{(\mathcal{M}_T)}\rsk\stackrel{d}{\longrightarrow}\mathcal{N}\lsk\boldsymbol {0}_{p_0},\lsk1+\frac{1}{\alpha}\rsk \lsk  V_{\bmbeta_0}^{-1}\Sigma_{\bmbeta_0}  V_{\bmbeta_0}^{-1}\rsk^{(\mathcal{M}_T)}\rsk.\]
 \end{theorem}

In Theorem \ref{tm:1}, (1) implies that the regularized subbagging estimator $\hbmbeta_{k_N,m_N}=({\hbmbeta_{k_N,m_N}^{(\mathcal{M}_T)}}^\top, $ ${\hbmbeta_{k_N,m_N}^{(-\mathcal{M}_T)}}^\top)^\top$ is $\sqrt{N}$-consistent. In addition, (2) states that the estimated coefficients of the irrelevant variables in $\bmx_i^{(-\mathcal{M}_T)}$ are zeros with probability tending to 1. Together, (1) and (2) yield that $\hbmbeta_{k_N,m_N}$ can identify the true model $\mathcal{M}_T$ consistently, i.e., the 
variable selection consistency. 
Moreover, (3) indicates that the estimated coefficients in $\hbmbeta_{k_N,m_N}^{(\mathcal{M}_T)}$ that correspond to the relevant variables in $\bmx_i^{(\mathcal{M}_T)}$ are asymptotically normal, which allows for statistical inference, such as hypothesis testing and confidence interval construction. Finally, 
(3) suggests that $\hbmbeta_{k_N,m_N}^{(\mathcal{M}_T)}$ enjoys the oracle properties (see, e.g., \citealp{wang2007unified}) in the sense that the coefficients of the relevant variables in $\bmx_i^{(\mathcal{M}_T)}$ can be estimated as efficiently as if the true model $\mathcal{M}_T$ were known in advance.


We next compare our estimator $\hbmbeta_{k_N,m_N}$ to 
the full sample regularized estimator $\hbmbeta_N$ which minimizes
 (\ref{global:al}). Using the similar techniques to those used in the proof of Theorem \ref{tm:1} in Section \ref{subsec: thproof} of the supplementary material, one can show that 
 \be
 \sqrt{N}\lsk\hbmbeta_{N}^{(\mathcal{M}_T)}-\bmbeta_0^{(\mathcal{M}_T)}\rsk\stackrel{d}{\longrightarrow}\mathcal{N}\lsk\boldsymbol {0}_{p_0}, \lsk  V_{\bmbeta_0}^{-1}\Sigma_{\bmbeta_0} V_{\bmbeta_0}^{-1}\rsk^{(\mathcal{M}_T)}\rsk.
 \label{eq:lasso}
 \ee
Accordingly, even though the regularized subbagging estimator $\hbmbeta_{k_N,m_N}^{(\mathcal{M}_T)}$ is also $\sqrt N$-consistent, it incurs a variance inflation rate of $1+1/\alpha$ compared to 
 $\hbmbeta_N^{(\mathcal{M}_T)}$. To remove this variance inflation, we can consider $k_Nm_N/N\to\alpha=\infty$ and define $1/\alpha = 0$, which then yields $1+1/\alpha=1$. In this case, the number of subsamples $m_N$ must grow faster than $N/k_N$, resulting in a relatively higher computational cost but achieving the same asymptotic efficiency as the full sample regularized estimator $\hbmbeta_N^{(\mathcal{M}_T)}$. This variance inflation for $\alpha<\infty$ arises from the potential overlap between any two subsamples in $\mS = \{s^{(1)}, \cdots, s^{(m_N)}\}$, where each subsample $s^{(\ell)}$ is independently drawn from the full sample $\{1,\cdots,N\}$ of $\ell=1,\cdots,m_N$. More discussion of this variance inflation and its related theory can be found in \cite{zou2021subbagging}.





In this paper, we primarily focus on achieving the convergence rate of $O_{\rmP}(N^{-1/2})$  for the regularized subbagging estimator, i.e., the same rate as that for the full sample estimation, and hence we consider the setting of $k_Nm_N/N\to\alpha\in(0,\infty]$. As $\alpha=0$, using techniques similar to those in the proof of Theorem \ref{tm:1}, we can show that $\hbmbeta_{k_N,m_N} -\boldsymbol{ \beta}_{0}=O_{\rmP}\big(({k_Nm_N})^{-1/2}\big)$, which converges more slowly than $O_{\rmP}(N^{-1/2})$ when $k_Nm_N/N\to\alpha=0$. Hence, we do not consider $\alpha=0$ in the rest of this paper.




Theorem \ref{tm:1} imposes the conditions $k_N/N\to 0$ and $k_N/\sqrt{N}\to\infty$, indicating that the subsample size $k_N$ can be chosen to be small but must have an order of $k_N\asymp N^{1/2+\delta}$ for some $0<\delta<1/2$. The reason for this is provided below.  Specifically, the regularized subbagging estimator $\hbmbeta_{k_N,m_N}$ in (\ref{loss:reg}) can be treated as an aggregation of the subsample estimators $\tbmbeta_{k_N,s}$ for $s\in\mS$, where Lemma \ref{lm:5} of the supplementary material shows that the bias of $\tbmbeta_{k_N,s}$ is of order $O(1/k_N)$. Aggregating 
$\tbmbeta_{k_N,s}$ can reduce the variance to the order of $O\big((1+1/\alpha)N^{-1}\big)$, but does not improved the order of the bias, which remains at $O(1/k_N)$ (see  \citealp{zou2021subbagging}). As a  consequence, the aggregation of $\tbmbeta_{k_N,s}$ has a mean squared error (MSE) of order $O(N^{-1}+k_N^{-2})=O(N^{-1})$ only when $k_N/\sqrt{N}\to\infty$. This is the intuitive reason why the regularized subbagging estimator $\hbmbeta_{k_N,m_N}$ can achieve $\sqrt N$-consistency in Theorem \ref{tm:1}; details are provide in the proof of Theorem \ref{tm:1}.

To satisfy the requirement of $k_N$ above, we select $k_N\asymp N^{1/2+\delta}$ with $\delta = 1/4$ and $1/3$  in simulation studies of Sections \ref{sec:simu}, and $\delta = 1/4$ in real data analysis of Section \ref{sec:real}. This, together with the condition $k_Nm_N/N\to\alpha\in(0,\infty]$ required in Theorem \ref{tm:1}, leads to the selection of $m_N\asymp N^{1/2-\delta+\delta'}$, where $\delta' = 0$ corresponds to the minimum requirement of $m_N$ but results in a variance inflation of $1+1/\alpha$, while $\delta'>0$ eliminates variance inflation at the cost of increased computation of more subsamples. In our numerical studies of Sections \ref{sec:simu} -- \ref{sec:real}, we select $\delta'=0$.

 




Recall that (3) in Theorem \ref{tm:1} allows for statistical inference. 
Specifically, let $\widehat{\mathcal{M}}_T=\{j:\hat\beta_{k_N,m_N,j}\neq 0\}$. Note that $k_N/N\to 0$ and $k_Nm_N/N\to\alpha$ in Theorem \ref{tm:1} implies $m_N\to\infty$. This allows us to 
adopt the subbagging variance estimation approach proposed in \cite{zou2021subbagging}, i.e., 
\be\label{eq:omega1}
\widehat\Psi_{k_N,m_N}=\frac{k_N}{m_N}\sum_{s\in\mS}\lsk \hbmbeta_{k_N,s}^{(\widehat{\mathcal{M}}_T)}- \hbmbeta_{k_N,m_N}^{(\widehat{\mathcal{M}}_T)}\rsk \lsk \hbmbeta_{k_N,s}^{(\widehat{\mathcal{M}}_T)}- \hbmbeta_{k_N,m_N}^{(\widehat{\mathcal{M}}_T)}\rsk^\top,\ee
which can be used to estimate $(  V_{\bmbeta_0}^{-1}\Sigma_{\bmbeta_0}  V_{\bmbeta_0}^{-1})^{(\mathcal{M}_T)}$ in (3). 
This estimator is obtained based on the asymptotic normality of the subsample estimator $\tbmbeta_{k_N,s}$, i.e.,
$k_N^{1/2}(\tbmbeta_{k_N,s}-\bmbeta_0) \stackrel{d}{\longrightarrow} \mathcal{N}(\bmo_{p},V_{\bmbeta_0}^{-1}\Sigma_{\bmbeta_0}  V_{\bmbeta_0}^{-1})$ in Lemma \ref{lm:5} of the supplementary material,
and is analogous to the bootstrap variance estimator, except that it is based on drawing $m_N$ subsamples rather than $m_N$ bootstrap samples. Using  (3) in Theorem \ref{tm:1} and the variance estimator (\ref{eq:omega1}), one can test hypotheses and construct confidence intervals for $\bmbeta^{(\mathcal{M}_T)}$.






In addition, we recall that $a_w = \max\{\lambda w_j :1\leq j \leq p_0\}$ and $b_w = \min\{\lambda w_j :p_0<j \leq p\}$. The condition $\sqrt{N}a_w\stackrel{\rmP}{\longrightarrow} 0$ imposed in (1) of Theorem \ref{tm:1}  
indicates that the penalty on the nonzero coefficients of $\bmbeta$ needs to be small, while $\sqrt{N}b_w\stackrel{\rmP}{\longrightarrow} \infty$ imposed in (2) of Theorem \ref{tm:1} implies that the penalty on the zero coefficients of $\bmbeta$ must be large enough to achieve variable selection consistency. Note that in this paper, we choose  $w_j=1/|\tilde\beta_{k_N,m_N,j}|^\gamma$ below (\ref{loss:reg}), where Lemma \ref{ps:3} of the supplementary material shows that $\sqrt{N}(\tilde\beta_{k_N,m_N,j}-\beta_{0,j})$ converges in distribution to a normal random variable. 
Accordingly,  $w_j=O_\rmP(1)$ for $j=1,\cdots, p_0$, and $w_j\asymp N^{\gamma/2}$ in probability for $j=(p_0+1),\cdots, p$; this is because the first $p_0$ elements $\beta_{0,1},\cdots,\beta_{0,p_0}$ are nonzero, while the remaining  $p-p_0$ elements $\beta_{0,p_0+1},\cdots,\beta_{0,p}$ are zeros. Under this setting of $w_j$, we can theoretically choose $\lambda\asymp (\log N)/N$, which guarantees that both $\sqrt{N}a_w\stackrel{\rmP}{\longrightarrow} 0$ and $\sqrt{N}b_w\stackrel{\rmP}{\longrightarrow} \infty$ hold if we set $\gamma\geq 1$. On the other hand, we will also provide a data-driven approach to select the regularizaiton parameter $\lambda$ in the next section.

\subsection{Subbagging Bayesian Information Criterion}\label{sec:SBIC}

A significant portion of the literature adapts traditional model selection criteria, such as the Akaike information criterion (AIC,  \citealp{akaike1973maximum}) and the Bayesian information criterion (BIC,  \citealp{schwarz1978estimating}), to develop data-driven methods for selecting the regularization parameter $\lambda$. In addition, the generalized cross-validation approach \citep{golub1979generalized} is also commonly used for selecting $\lambda$. However, \citet{wang2007tuning} and \citet{zhang2010regularization} noted that both AIC and generalized cross-validation for selecting $\lambda$ achieve only screening consistency (i.e., no false exclusion) but do not guarantee variable selection consistency. To this end, we adapt the BIC-type criterion from \citet{wang2007tuning} and \citet{zhu2021least} to propose the subbagging Bayesian information criterion (SBIC) for selecting $\lambda$ given below.

Specifically, we slightly abuse the notation and denote the regularized subbagging estimator $\hbmbeta_{k_N,m_N}$ as $\hbmbeta_{k_N,m_N,\lambda}$ in this section; this is because  $\hbmbeta_{k_N,m_N}$ is computed using (\ref{loss:reg}), which varies with different values of $\lambda$. In addition, let $\text{DF}_{\lambda}$ be the number of nonzero elements in  $\hbmbeta_{k_N,m_N,\lambda}\in\mathbb{R}^p$. Then the proposed SBIC is defined as follows: 
\be\label{SBIC:def}\mathrm{\text{SBIC}}_{\lambda}={k_N}\tmL_{k_N,m_N}\lsk\hbmbeta_{k_N,m_N,\lambda}\rsk+{\log  N}  \times \text{DF}_{\lambda}.
\ee
Subsequently, the regularization parameter $\lambda$ can be selected by  $\hat\lambda=\arg\min_\lambda\mathrm{\text{SBIC}}_{\lambda}$. Similar to the interpretation of the classical BIC, the measure of goodness of fit in (\ref{SBIC:def}) is given by the subbagging loss function evaluated at $\hbmbeta_{k_N,m_N,\lambda}$, i.e., $\tmL_{k_N,m_N}(\hbmbeta_{k_N,m_N,\lambda})$, scaled by $k_N$, while the measure of model complexity in (\ref{SBIC:def}) is represented by $\text{DF}_{\lambda}$. The rationale for using the scaling factor $k_N$ before the subbagging loss function will be explained in Remark \ref{rm:BIC} below.

Recall that before (\ref{eq:omega1}), we denote the estimated model by $\widehat{\mathcal{M}}_T=\{j:\hat\beta_{k_N,m_N, j}\neq 0\}$. Using the notation $\hbmbeta_{k_N,m_N,\lambda}$ in this section, we redefine the estimated model by 
$\widehat{\mathcal{M}}_\lambda=\{j:\hat\beta_{k_N,m_N,\lambda, j}\neq 0\}$, 
where $\hat\beta_{k_N,m_N,\lambda, j}$ is the $j$-th element of $\hbmbeta_{k_N,m_N,\lambda}$. Then, we can partition $\lambda$ into  three mutually exclusive sets $\Omega_{0}=\{\lambda \in \mathbb{R}: \widehat{\mathcal{M}}_{\lambda}=\mathcal{M}_{T}\}$, $\Omega_{-}=\{\lambda \in \mathbb{R}: \widehat{\mathcal{M}}_{\lambda} \not \supset \mathcal{M}_{T}\}$ and  $\Omega_{+}=\{\lambda \in \mathbb{R}: \widehat{\mathcal{M}}_{\lambda} \supset$ $\mathcal{M}_{T}\textrm{ and }\widehat{\mathcal{M}}_{\lambda} \neq \mathcal{M}_{T}\} .$ 
We next establish the theoretical guarantee for the SBIC given below.


 \begin{theorem}\label{tm:BIC}Under Conditions  {(C1) -- (C5)} in Appendix, assume $k_N\to\infty$, $k_N/N\to 0$, $k_N/\sqrt{N}\to\infty$ and $k_Nm_N/N\to\alpha\in(0,\infty]$  as $N\to \infty$.  For any given reference regularization parameter $\lambda_N$ satisfying $\lambda_N\asymp (\log N)/N$, we have
\[\rmP\left(\inf _{\lambda \in \Omega_{-} \cup \Omega_{+}} \mathrm{SBIC}_{\lambda}>\mathrm{SBIC}_{\lambda_{N}}\right) \rightarrow 1.\]
 \end{theorem}

Based on Theorem \ref{tm:BIC}, we next explain why the regularization $\lambda$ selected by the SBIC can achieve variable section consistency. At the end of Section \ref{subsec:AT}, we have mentioned that the reference regularization parameter $\lambda_N\asymp (\log N) /N$ guarantees both  $\sqrt{N}a_w\stackrel{\rmP}{\longrightarrow} 0$ and $\sqrt{N}b_w\stackrel{\rmP}{\longrightarrow} \infty$ in Theorem \ref{tm:1}, if we set $\gamma\geq 1$ in the adaptive weight $w_j=1/|\tilde\beta_{k_N,m_N,j}|^\gamma$. Hence, $\lambda_N$ ensures variable selection consistency in Theorem \ref{tm:1}, meaning that the estimated model $\widehat{\mathcal{M}}_{\lambda_N}=\mathcal{M}_T$ with probability tending to 1. This is why we refer to $\lambda_N$ as the reference regularization parameter.
On the other hand, Theorem \ref{tm:BIC} indicates that if $\lambda$ falls within the sets $\Omega_{-}$ or $\Omega_{+}$ (either of which fails to identify the true model $\mathcal{M}_T$), then the corresponding $\mathrm{SBIC}_{\lambda}$ will be greater than $\mathrm{SBIC}_{\lambda_{N}}$ with probability tending to 1.
 For $\hat{\lambda}$ that minimizes $\mathrm{SBIC}_{\lambda}$, we have $\mathrm{SBIC}_{\hat \lambda} \leq \mathrm{SBIC}_{\lambda_{N}}$. Hence, the only possibility is that $\hat{\lambda}$ falls within $\Omega_{0} = \{\lambda \in \mathbb{R}: \widehat{\mathcal{M}}_{\lambda} = \mathcal{M}_{T} \}$, thereby ensuring variable selection consistency.


\begin{remark}\label{rm:BIC}
\normalfont
By Lemma \ref{lm:BIC2} of the supplementary material, the subbagging loss function $\tmL_{k_N,m_N}(\bmbeta)$ evaluated at any $\sqrt N$-consistent estimator $\hbmbeta$ is of order $O_\rmP(k_N^{-1})$. As a consequence, we scale $\tmL_{k_N,m_N}(\bmbeta)$ by $k_N$ in the SBIC (\ref{SBIC:def}). This scaling differs from the existing literature on BIC regularization parameter selection (see, e.g., \citealp{wang2007unified} and \citealp{zhu2021least}), and it is crucial for establishing the variable selection consistency result in Theorem \ref{tm:BIC}; see the proof of Theorem \ref{tm:BIC} in Section \ref{subsec: thproof} of the supplementary material.
\end{remark}

\section{Simulation Studies}
\label{sec:simu}

In this section, we evaluate the numerical performance of the regularized subbagging estimator $\hbmbeta_{k_N,m_N}$ in (\ref{loss:reg}),   and compare it with the full sample regularized estimator $\hbmbeta_N$ which minimizes
 (\ref{global:al}).  All simulation results, except Tables \ref{tb:2} -- \ref{tb:4}, are provided in Section \ref{sec:ASimu} of the supplementary material to save space. 
 
 In our simulation, the full sample $\{\bmz_i=(y_i,\bmx_i^\top)^\top:i=1,\cdots,N\}$ is simulated from logistic and linear regression models, where $y_i$ is the response and $\bmx_i=(x_{i,1},\cdots,x_{i,8})^\top$ is the covariate vector. For logistic regression, the binary response $y_i$ is independently generated from a Bernoulli distribution with 
$
\rmP \lsk y_{i}=1\vert \boldsymbol x_{i}\rsk={\exp\lsk\boldsymbol x_{i}^{\top} \bmbeta_{0}\rsk}/\{1+\exp\lsk\boldsymbol x_{i}^{\top} \bmbeta_{0}\rsk\}. 
$
For linear regression, $y_i$ is generated from $y_{i}=\boldsymbol x_{i}^{\top} \bmbeta_{0}+\varepsilon_{i},$ 
where  $\varepsilon_{i}$
is independently drawn from the standard normal distribution $\mathcal{N}(0,1)$. Following \cite{hunter2005variable},  we set $\bmbeta_0=(\beta_{0,1},\cdots,\beta_{0,8})^\top= (3, 1.5, 2, 0, 0, 0, 0, 0)^\top$ for both models, and hence the true model $\mathcal{M}_T=\{1,2,3\}$.  The covariates $x_{i,1},\cdots,x_{i,8}$ are independently generated from $\mathcal{N}(0,1)$.

To simulate the environment of big data, we consider $N= 500,000$ and $1,000,000$. To obtain the regularized subbagging estimate $\hbmbeta_{k_N,m_N}$, we select two subsample sizes,  $k_N=\lfloor N^{1/2+1/4}\rfloor$ and $\lfloor N^{1/2+1/3}\rfloor$, where $\lfloor\cdot\rfloor$ denotes the floor  function, and we set the number of subsamples by $m_N=\lfloor \alpha N/k_N\rfloor$ for $\alpha =0.1$, $0.5$ and $1$. These settings satisfy the conditions in Theorem \ref{tm:1}. The regularization parameter $\lambda$ used to compute $\hbmbeta_{k_N,m_N}$ is selected by $\hat\lambda=\arg\min_\lambda\mathrm{\text{SBIC}}_{\lambda}$, where this minimization is carried out using a one-dimensional log-scale grid search,  similar in principle to the approach in \cite{fan2017estimation}. Analogously, to compute the full sample regularized estimator $\hbmbeta_N$, $\lambda$ is selected based on the BIC-type criterion from \citet{wang2007tuning}.



For each of the above simulation settings, we obtain $r=1,\cdots,$ 1,000 replications   $\hbmbeta^{(r)}=(\hat\beta_1^{(r)},\cdots,\hat\beta_8^{(r)})^\top$ of estimate $\hbmbeta=(\hat\beta_1,\cdots,\hat\beta_8)^\top$, where $\hbmbeta$ represents  either $\hbmbeta_{k_N,m_N}$ or $\hbmbeta_N$. Using these replications, we evaluate the estimation performance of $\hbmbeta^{(\mathcal{M}_T)}$ for $\mathcal{M}_T=\{1,2,3\}$, as well as the variable selection performance based on $\hbmbeta$.

To assess the estimation performance, we consider five measures of estimate $\hat\beta_j$ for  $j\in\mathcal{M}_T$. The first three are: the empirical bias (BIAS)  $1000^{-1}\sum_{r=1}^{1000}( \hat{\beta}_{j}^{(r)}-{\beta}_{0,j})$, the empirical standard deviation (SD) $\{1000^{-1}\sum_{r=1}^{1000}(\hat{\beta}_{j}^{(r)}-1000^{-1}{\sum_{r=1}^{1000}\hat{\beta}_{j}^{(r)}})^2\}^{1/2}$, and the root mean squared error (RMSE) $(\textrm{BIAS}^2+\textrm{SD}^2)^{1/2}$. The fourth measure is the averaged standard error (ASE) $1000^{-1}\sum_{r=1}^{1000}\textrm{SE}_j^{(r)}$, where $\textrm{SE}_j^{(r)}$ is the $r$-th replication of $\hat\beta_j$'s standard error $\textrm{SE}_j$. For the regularized subbagging estimator $\hbmbeta_{k_N,m_N}$, $\textrm{SE}_j$ is obtained by $[N^{-1}\{1+N/(k_Nm_N)\}(\widehat\Psi_{k_N,m_N})_{j,j}]^{1/2}$, where $(\widehat\Psi_{k_N,m_N})_{j,j}$ denotes the $(j,j)$-th element of $\widehat\Psi_{k_N,m_N}$, and  $\widehat\Psi_{k_N,m_N}$ is   (\ref{eq:omega1}) by letting $\widehat{\mathcal{M}}_T=\mathcal{M}_T$. For the full sample regularized estimator $\hbmbeta_N$, $\textrm{SE}_j$ is obtained by $\{N^{-1}(\widehat \Psi_{N})_{j,j}\}^{1/2}$, where $\widehat\Psi_{N}=(\widehat{V}_{\hbmbeta_N}^{-1}\widehat{\Sigma}_{\hbmbeta_N}\widehat{V}_{\hbmbeta_N}^{-1})^{(\mathcal{M}_T)}$ with 
$\widehat{\Sigma}_{\hbmbeta_N}=N^{-1}\sum_{i=1}^{N}\{$ $\nabla \mL(\hbmbeta_N;\bmz_i)-N^{-1}\sum_{i=1}^{N}\nabla \mL(\hbmbeta_N;\bmz_i)\}\{\nabla \mL(\hbmbeta_N;\bmz_i)-N^{-1}\sum_{i=1}^{N}\nabla \mL(\hbmbeta_N;\bmz_i)\}^\top$ 
and $\widehat{V}_{\hbmbeta_N}=N^{-1}\sum_{i=1}^{N}\nabla^2 \mL(\hbmbeta_N;\bmz_i)$.
The fifth measure is 
 the empirical coverage probability (CP) for the 95\% confidence interval constructed from $\hat\beta_j$ and its corresponding asymptotic distribution. Specifically, the CP is
 $1000^{-1} \sum_{r=1}^{1000} \mathbbm{1}\{
 \beta_{0,j}  \in [\hat\beta_j^{(r)}-1.96\times \textrm{SE}_j^{(r)},\hat\beta_j^{(r)}+1.96\times \textrm{SE}_j^{(r)}]
 \}$, where $\mathbbm{1}\{\cdot\}$ is the indicator function.

To assess the variable selection performance, we denote the estimated model based on  $\hbmbeta$ as $\widehat{\mathcal{M}}=\{j:\hat\beta_j\neq 0\}$. Since we obtain $r=1,\cdots,$ 1,000 replications   $\hbmbeta^{(r)}$ of $\hbmbeta$, we also have 1,000 replications $\widehat{\mathcal{M}}^{(r)}$ of the estimated model $\widehat{\mathcal{M}}$. 
We subsequently consider the  five measures of $\widehat{\mathcal{M}}$: (i) the averaged correct fit (CF) $1000^{-1}\sum_{r=1}^{1000}\mathbbm{1}\{\widehat{\mathcal{M}}^{(r)}=\mathcal{M}_{T}\}$;  (ii)  the
averaged true positive rate (TP) $1000^{-1}\sum_{r=1}^{1000}|\widehat{\mathcal{M}}^{(r)} \cap \mathcal{M}_{T}| /|\mathcal{M}_{T}|$, where  $|\cdot|$ is the cardinality of any generic set; (iii) the averaged false positive rate (FP) $1000^{-1}\sum_{r=1}^{1000}|\widehat{\mathcal{M}}^{(r)}$ $\cap \mathcal{M}^{c}_{T}| /|\mathcal{M}^{c}_{T}|,$ where $\mathcal{M}^{c}_{T}=\mathcal{M}_F\backslash\mathcal{M}_T=\{4,5,6,7,8\}$; (iv) the averaged model size  (MS) $1000^{-1}\sum_{r=1}^{1000}|\widehat{\mathcal{M}}^{(r)}|$; and (v) the empirical standard deviation of model sizes (sd) $\{1000^{-1}$ $\sum_{r=1}^{1000}(|\widehat{\mathcal{M}}^{(r)}|-\textrm{MS})^2\}^{1/2}$.

Table \ref{tb:2} presents the five measures used to assess the estimation performance for logistic regression. To save space, Table \ref{tb:2} only presents the results for  \( k_N = \lfloor N^{1/2+1/4} \rfloor \), whereas Table \ref{tb:2:1} provides results for \( k_N = \lfloor N^{1/2+1/3} \rfloor \). Tables \ref{tb:2} and \ref{tb:2:1} reveal 
 four interesting findings.
First, as $N$ increases from 500,000 to 1,000,000, the RMSE of the regularized subbagging estimate $\hbmbeta_{k_N,m_N}$ decreases, which empirically supports its consistency as $N\to\infty$. 
Second, for a given $N$, $\hbmbeta_{k_N,m_N}$ exhibits a larger BIAS compared to the regularized full sample estimate $\hbmbeta_{N}$, while increasing $k_N$ reduces its BIAS.
 Third, for a given $N$ and $k_N$, increasing $m_N$ (by setting a larger $\alpha$) reduces the SD and RMSE of $\hbmbeta_{k_N,m_N}$  (bringing them closer to those of $\hbmbeta_{N}$), but does not decrease the BIAS for $\hbmbeta_{k_N,m_N}$.
 The second and third findings align with the comment after Theorem \ref{tm:1}, stating that the subbagging procedure reduces the variance to $O\big((1+1/\alpha)N^{-1}\big)$ but does not improve the bias, which remains at an order of $O(k_N^{-1})$. 
 Lastly, the CP constructed from $\hbmbeta_{k_N,m_N}$ is close to 95\% across all settings, confirming the asymptotic normality in (3) of Theorem \ref{tm:1} and the validity of the subbagging variance estimator in (\ref{eq:omega1}).



\begin{table}[htpb!]
\renewcommand{\arraystretch}{0.8}
\setlength{\tabcolsep}{5pt}
\caption{The five measures used to assess the estimation performance under the settings of logistic regression and $k_N=\lfloor N^{1/2+1/4}\rfloor$. The BIAS, SD, RMSE and ASE  in this table are 100 times  their actual values.} \label{tb:2}
\vspace{-20 pt}
\begin{center}
 \setlength\extrarowheight{-3pt}
\scalebox{0.9}{
\begin{tabular}{c|c|rrr|rrr|rrr|rrr}
   \hline

 &\multicolumn{1}{c|}{}&\multicolumn{9}{c}{{$\hbmbeta_{k_N,m_N}$ with $m_N=\lfloor \alpha N/k_N\rfloor$}} &\multicolumn{3}{|c}{$\hbmbeta_{N}$}   \\
\cline{3-4}\cline{5-6}\cline{7-8}\cline{9-11}\cline{12-14}
$N$& \multicolumn{1}{c|}{}& \multicolumn{1}{c}{$\hat\beta_1$}& \multicolumn{1}{c}{$\hat\beta_2$} &  \multicolumn{1}{c|}{$\hat\beta_3$} & \multicolumn{1}{c}{$\hat\beta_1$}& \multicolumn{1}{c}{$\hat\beta_2$} &  \multicolumn{1}{c|}{$\hat\beta_3$} & \multicolumn{1}{c}{$\hat\beta_1$}& \multicolumn{1}{c}{$\hat\beta_2$} &  \multicolumn{1}{c|}{$\hat\beta_3$} & \multicolumn{1}{c}{$\hat\beta_1$}& \multicolumn{1}{c}{$\hat\beta_2$} &  \multicolumn{1}{c}{$\hat\beta_3$} 
 
\\
\cline{3-4}\cline{5-6}\cline{7-8}\cline{9-11}\cline{12-14}
   && \multicolumn{3}{c|}{$ \alpha=0.1$}    & \multicolumn{3}{c|}{$\alpha=0.5$}      &
\multicolumn{3}{c|}{$ {\alpha=1}$} &\multicolumn{3}{c}{}
 \\
  \cline{1-2} \cline{3-4}\cline{5-6}\cline{7-8}\cline{9-11}\cline{12-14}

&BIAS&0.47 &0.21 &0.55&0.46 &0.23& 0.55&0.42 &0.22 &0.55&0.02 &0.01&-0.03\\
 & SD& 2.78& 1.85 &2.14&1.38 &0.95& 1.29&1.13& 0.80& 1.03&0.83&0.57&0.74 \\
 500,000&  RMSE &2.78 & 1.85& 2.20&1.45 &0.98& 1.40&1.20& 0.82& 1.23&0.83&0.57&0.74\\
&ASE&2.83&1.82&2.22&1.38&1.00&1.25&1.15&0.81&1.05&0.84&0.56&0.73\\
& CP (\%)& 94.3&93.2&94.1&93.4&93.3&93.1&93.2&93.6&93.7&95.0& 93.2&93.5\\
\cline{1-4}\cline{5-6}\cline{7-8}\cline{9-14}
&BIAS&0.01 &-0.17&  0.01&0.02&-0.16&-0.05&-0.01 &-0.17 &-0.04&-0.00&0.01&-0.02\\
 & SD&1.88& 1.21& 1.47&0.98& 0.60&  0.81&  0.69& 0.43 &0.63&0.58&0.39&0.51  \\
1,000,000&  RMSE &1.88& 1.22& 1.47&0.98& 0.62 &0.82&0.69 &0.46&  0.63&0.58&0.39&0.51 \\
&ASE&1.85&1.22&1.48&0.96&0.61&0.81&0.70&0.45&0.61&0.58&0.39&0.51\\
& CP (\%)&93.2&96.2&94.6&93.0&94.8&94.3&95.1&94.3&94.6&95.8&93.9&96.1\\

\hline
 \end{tabular}}
\end{center}
\vspace{-20pt}
\end{table}

Recall that the regularization parameter $\lambda$ utilized to compute $\hbmbeta_{k_N,m_N}$ is selected by $\hat\lambda=\arg\min_\lambda\mathrm{\text{SBIC}}_{\lambda}$.  Table \ref{tb:4} reports the five measures used to assess the variable selection performance based on  $\hbmbeta_{k_N,m_N}$  for logistic regression. To save space, Table \ref{tb:4} only presents the results for  \( k_N = \lfloor N^{1/2+1/4} \rfloor \), whereas Table \ref{tb:4:1} provides results for \( k_N = \lfloor N^{1/2+1/3} \rfloor \).  
Tables \ref{tb:4} and \ref{tb:4:1} reveal two insights. 
First, using $\textrm{SBIC}_\lambda$ to obtain $\hbmbeta_{k_N,m_N}$ consistently selects three variables in MZ, which matches the true model size, $|\mathcal{M}_T|=3$.  
Second, applying $\textrm{SBIC}_\lambda$ to obtain $\hbmbeta_{k_N,m_N}$ results in 100\% CF, 100\% TP, and 0\% FP across all settings, even for smaller values of $m_N$ and $N$.  
These results demonstrate the variable selection consistency achieved by $\textrm{SBIC}_\lambda$.

\begin{table}[htpb!]
\renewcommand{\arraystretch}{0.8}
\setlength{\tabcolsep}{2pt}
\caption{The five measures used to assess the variable selection performance based on  $\hbmbeta_{k_N,m_N}$ under the settings of logistic regression and $k_N = \lfloor N^{1/2+1/4} \rfloor$.  } \label{tb:4}
\vspace{-20pt}
\begin{center}
 \setlength\extrarowheight{-3pt}
\scalebox{1}{
\begin{tabular}{cc|r|r|r||r|r|r}
   \hline

\multicolumn{8}{c}{$m_N=\lfloor \alpha N/k_N\rfloor$}   \\
\hline
&\multirow{2}{*}{$\quad$}&\multicolumn{3}{c||}{$N=$ 500,000}&\multicolumn{3}{c}{$N=$ 1,000,000}   \\
\cline{3-8}
 && \multicolumn{1}{c|}{$ \alpha=0.1$}    & \multicolumn{1}{c|}{$ \alpha=0.5$}      &
\multicolumn{1}{c||}{$ \alpha=1$}  & \multicolumn{1}{c|}{$ \alpha=0.1$}    & \multicolumn{1}{c|}{$ \alpha=0.5$}      &
\multicolumn{1}{c}{$ \alpha=1$} 
\\
\cline{3-4}\cline{5-6}\cline{7-8}
 
 \cline{3-4}\cline{5-6}\cline{7-8}
\hline
&CF (\%)&100.00&100.00&100.00&100.00&100.00&100.00                          \\
&TP (\%)&100.00&100.00&100.00&100.00&100.00&100.00\\
&FP (\%)&0.00& 0.00&0.00     &0.00& 0.00&0.00                          \\
&MZ (sd)&3.00 (0.00)   &3.00 (0.00) &3.00 (0.00)&3.00 (0.00)   &3.00 (0.00) &3.00 (0.00)       \\
\hline


 

\end{tabular}}
\end{center}
\vspace{-24pt}
\end{table}

Tables \ref{tb:1} -- \ref{tb:3} report the results for linear regression. These tables show qualitatively similar findings to those in Tables \ref{tb:2} and \ref{tb:2:1} for the estimation performance, and to those in Tables \ref{tb:4} and \ref{tb:4:1} for the variable selection performance. Additional findings related to these tables are provided in Section \ref{sup:Asimfind} of the supplementary material.



\section{Real Data Analysis}
\label{sec:real}

In this section, we apply our proposed method to analyze a public dataset known as Public Use Microdata Areas (PUMAs) (\url{https://www.census.gov/programs-surveys/geography/guidance/geo-areas/pumas.html}), released by the U.S. Census Bureau. The dataset contains individual resident records and their associated characteristics (see, e.g.,  \citealp{gbur2004statistical} for details), providing fundamental information for studying socio-economic issues. In this study, we aim to identify key factors that influence the high-income status of individuals, and estimate their associated effects.

Specifically, the dataset utilized in this  study comprises the census information of PUMAs spanning from 2014 to 2018, with a total of $N=15,965,200$ observations in the full sample. For each observation $(y_i,\bmx_i^\top)^\top$, $i=1,\cdots,N$, the response variable $y_i=1$ if the individual's annual income surpasses the high-income threshold \$67,420 (twice the U.S. median income to account for labor productivity growth and inflation, suggested by \citealp{kochhar2018american}); otherwise $y_i=0$. 
To predict the high-income status $y_i$, we construct the covariate vector $\bmx_i = (1, x_{i2}, \cdots, x_{ip})^\top$ for $i = 1, \cdots, N$, partly following \cite{kohavi1996scaling}. First, $x_{i2},\cdots, x_{i6}$ are ``Age'', ``Capital gain in an investment'', ``Capital loss in an investment'', ``The number of years of education before high school'' 
and ``Usual hours worked per week  past 12 months''. 
The rest of the covariates are indicator variables representing eight categorical variables:
``Ancestry'' (White, Asian-Pacific Islander, American Indian-Eskimo, Black, Other), 
``Citizenship status" (Born in the U.S., Born abroad of U.S. citizen parent or parents, U.S. citizen by naturalization, Not a citizen of the U.S.), 
``Class of worker" (Private, Self-employed, Federal government, Local government, State government, Without pay,  Never worked), ``Educational attainment'' (No diploma completed, Regular high school diploma, Associate's degree, Bachelor's degree, Master's degree, Doctorate degree), 
``Gender'' (Female, Male), 
``Marital status'' (Married, Widowed, Divorced, Separated, Never married, under 15 years old),  
``Occupation'' (Technical support, Craft repair,  Sales, Executive managerial, Professional specialty, Handlers and cleaners, Machine operation and inspection, Administrative clerical, Farming and fishing, Transportation and moving, Private household services, Protective services, Armed forces, Other services),
and  ``Subfamily relationship'' (Wife, Own-child, Husband, Not-in-family, Unmarried, Other-relative). 
For each categorical variable, we construct its associated indicator variables by selecting the first level in the parentheses above as the reference level.





To demonstrate our proposed method, we fit a logistic regression model relating $y_i$ to $\bmx_i$, and
obtain the regularized subbagging estimate $\hbmbeta_{k_N,m_N}$ via
setting the subsample size as $k_N=\lfloor N^{1/2+1/4}\rfloor=252,569$, and the number of subsamples as $m_N=\lfloor\alpha N/k_N\rfloor=6$ and $30$, which correspond to $\alpha=0.1$ and $0.5$, respectively. These settings satisfy the conditions in Theorem \ref{tm:1}. The regularization parameter $\lambda$ used to compute $\hbmbeta_{k_N,m_N}$ is selected by $\hat\lambda=\arg\min_\lambda\mathrm{\text{SBIC}}_{\lambda}$. 
Table \ref{tb:5} presents the variables selected based on $\hbmbeta_{k_N,m_N}$, namely $\{j:\hat\beta_{k_N,m_N,j}\neq 0\}$, along with their associated coefficient estimates, standard errors (SEs), and $p$-values, where the SE is computed using the same approach as that in Section \ref{sec:simu}. Table \ref{tb:5:1} in the supplementary material reports the variables which are not selected based on $\hbmbeta_{k_N,m_N}$. Note that using different numbers of subsamples ($m_N = 6$ and 30) leads to the same variable selection outcome; see Tables \ref{tb:5} and \ref{tb:5:1}.

\begin{table}[htpb!]
\renewcommand{\arraystretch}{0.5}
\setlength{\tabcolsep}{2pt}
  \caption{The variables selected based on $\hbmbeta_{k_N,m_N}$, along with their associated coefficient estimates, SEs and $p$-values for the PUMAs data. The SEs are $10^{3}$ times their actual values.} \label{tb:5}
  \vspace{-18pt}
  \begin{center}
  \scalebox{1}{
  \begin{tabular}{ll|crl|crlcrlrrr}
     \hline

  \cline{3-4}\cline{5-6}\cline{7-8}\cline{9-11}\cline{12-14}
   \multicolumn{2}{c|}{\multirow{2}{*}{Variable Selected}}   & \multicolumn{3}{c|}{$m_N=6$ ($\alpha=0.1$)}    & \multicolumn{3}{c}{$m_N=30$ ($\alpha=0.5$)}      & 
   \\
    \cline{3-4}\cline{5-6}\cline{7-8}
        \cline{3-4}\cline{5-6}\cline{7-8}\cline{9-10}\cline{11-12}\cline{13-14}
    & \multicolumn{1}{c|}{}&  \multicolumn{1}{c}{estimate}& \multicolumn{1}{c}{SE}&\multicolumn{1}{c|}{$p$-value} &  \multicolumn{1}{c}{estimate}& \multicolumn{1}{c}{SE}&\multicolumn{1}{c}{$p$-value}&   \\\hline
    \multicolumn{2}{l|}{Intercept}& -9.078&44.315 &$<10^{-16}$&-9.067&17.037&$<10^{-16}$& \\
    \hline
    \multicolumn{2}{l|}{Age}& 0.027&0.022 &$<10^{-16}$& 0.024&0.005  &$<10^{-16}$& \\
    \hline
   \multicolumn{2}{l|}{Capital gain in  an investment} &0.003 &0.010 &$<10^{-16}$ &0.003 &0.002 &$<10^{-16}$ & \\ 
    \hline
\multicolumn{2}{l|}{\parbox{4.4cm}{\vspace{0.2em} Usual hours worked\vspace{-0.8em} \\ per week past 12 months\vspace{0.2em} }}&3.001& 11.223&$<10^{-16}$&2.987& 2.123&$<10^{-16}$& \\
    \hline
 \multirow{3}{*}{{\parbox{1.5cm}{Class of\\ Worker}}} & Self-employed&-0.492&66.425&  $<10^{-13}$&-0.495 &12.223&  $<10^{-16}$&\\
  & Federal government &-0.323&  55.413&  $<10^{-8}$&-0.333&  11.227&  $<10^{-16}$& \\
    &  Never worked&-0.535&  44.515&  $<10^{-16}$&-0.539&  8.375&  $<10^{-16}$& \\
        \hline
          \multirow{4}{*}{{\parbox{2.5cm}{Educational \\attainment}}}  &  Associate's degree            &0.527 & 44.802&$<10^{-16}$&0.530 &8.627 &$<10^{-16}$&\\
   &  Bachelor's degree      &1.697&33.249&$<10^{-16}$&$1.683$& 6.452  &$<10^{-16}$&\\
  & Master's degree &2.158&33.200&$<10^{-16}$&2.154& 6.076&$<10^{-16}$&\\
  & Doctorate degree& 2.755&60.940&$<10^{-16}$& 2.780& 12.243&$<10^{-16}$& \\
   \hline
Gender&Male& 1.193&22.174&$<10^{-16}$& 1.201&12.961   &$<10^{-16}$& \\ 
    \hline
Occupation&    Executive managerial&0.888&  88.513&$<10^{-16}$&0.890&  15.272 &$<10^{-16}$&\\
\hline

  \end{tabular}}
  \end{center}
  \vspace{-18pt}
  \end{table}


We next focus on interpreting the  results of the selected variables in Table \ref{tb:5}, which show seven interesting findings. 
First, it is unsurprising that senior employees tend to earn higher incomes than junior employees; this is because 
 age is often highly correlated with working experience. 
Second, individuals with higher capital gains are more likely to have higher incomes, possibly due to the positive contribution of their successful investments to their income.
Third, the more time individuals spend per week for working, the higher their likelihood of being high-income earners. 
Fourth, regarding ``Class of worker", federal government employees are likely to earn less than their private sector counterparts (reference level), aligning with the finding in \cite{makridis2021there}. 
Fifth, all levels of ``Educational attainment'' have a positive impact on earning high incomes compared to the reference level (No diploma completed). Notably, individuals with Master's and Doctorate degrees exhibit particularly large positive impacts, which is sensible since higher education levels often lead to better job opportunities. 
Sixth, the coefficient for ``Gender" implies that male workers are more likely to receive higher incomes than female workers, consistent with the findings in \cite{meara2020gender}.  
Seventh, concerning ``Occupation", executive managers are more likely to be high-income earners. This finding is also sensible because executives typically earn higher salaries.


It is worth noting that the full dataset with $N=15,965,200$ observations occupies approximately 8 GB on the hard drive.  Accordingly, on a lower-performance computer with only 8 GB of memory, the full dataset cannot be analyzed, as the computer might run out of memory and experience significantly slow computation. However, we can still obtain the regularized subbagging estimate $\hbmbeta_{k_N,m_N}$ and its associated variable selection results in Table \ref{tb:5}; this is because each subsample of size $k_N=\lfloor N^{1/2+1/4}\rfloor=252,569$ occupies only around 128 MB in the subbagging procedure.

For comparison, we compute the full sample regularized estimate $\hbmbeta_N$ which minimizes (\ref{global:al}), on a higher-performance computer with 32 GB of memory; the results are presented in Table \ref{tb:5:2} of the supplementary material. Based on this table, $\hbmbeta_N$ selects the same variables as those in Table \ref{tb:5}, and their associated coefficient estimates are  numerically close. In addition, the significance of the coefficients is also the same as in Table \ref{tb:5}. 
Table \ref{tb:6} further compares the computational resource usage for obtaining $\hbmbeta_{k_N,m_N}$ (on an 8 GB memory computer) and $\hbmbeta_{N}$ (on a 32 GB memory computer).  This table shows that both the computation time (including the loading time, estimation time, and the time for computing SEs) and the memory usage for obtaining the regularized subbagging estimate $\hbmbeta_{k_N,m_N}$, are much less than those for obtaining the full sample regularized estimate $\hbmbeta_{N}$. All of these findings demonstrate that our regularized subbagging estimation effectively tackles the challenges of large $N$ while maintaining the same variable selection and estimation accuracy as the full sample regularized estimation.

\begin{table}[!htbp]
\renewcommand{\arraystretch}{0.8}
\setlength{\tabcolsep}{2pt}
\caption{The computation resource usage for obtaining  $\hbmbeta_{k_N,m_N}$ and $\hbmbeta_{N}$.
}\label{tb:6}
  \vspace{-18pt}
\begin{center}
\scalebox{0.8}{\begin{tabular}{c|c|c|c|c}
\hline 
&   \multicolumn{1}{c|}{Settings}&{\makecell{Loading Time $\&$\\ Estimation Time (mins)}}&{\makecell{Time for Computing\\  SEs (mins)}}&Memory (GB)\\
\hline
    $\hbmbeta_{k_N,m_N}$ &$m_N=6$ ($\alpha=0.1$)&
    0.25 & $<$0.01 &$<$1 \\
    ($k_N=252,569$) & $m_N=30$ ($\alpha=0.5$)&
    1.42 & $<$0.01 &$<$1 \\
    \hline
$\hbmbeta_{N}$&$N=$$15,965,200$&
    14.79 & 4.78 &  7.99 \\
\hline 
\end{tabular}}
\end{center}
 \vspace{-18pt}
\end{table}

\section{Conclusion}

\label{sec:co}

In this article, we propose a novel regularized subbagging estimator  to perform simultaneous  variable selection and parameter estimation for massive datasets.
This estimator is demonstrated to achieve $\sqrt{N}$-consistency, the same convergence rate as the full sample estimation, while also attaining variable selection consistency and the asymptotic normality. 
Furthermore, we propose a new SBIC for selecting the regularization parameter, ensuring that the resulting regularized subbagging estimator exhibits selection consistency.
Our numerical studies reveal that the proposed method enhances computational scalability for large $N$ while maintaining the same variable selection and estimation accuracy as the full sample regularized estimation. To expand the application of our method, one could explore using the group LASSO regularizer instead of the adaptive LASSO regularizer considered in this paper. The group LASSO may be more suitable for selecting entire categorical variables with many levels rather than selecting the indicator variables constructed for different levels. We believe such an extension would further strengthen the usefulness of subbagging variable selection for big data.

\renewcommand{\theequation}{A.\arabic{equation}}
\setcounter{equation}{0}

\renewcommand{\thesubsection}{A.\arabic{subsection}}

\section*{Appendix: Technical Conditions} 
We first introduce the notation used in the technical conditions. Recall that $\nabla\mL(\bmbeta;\bmz_i)=\partial \mL(\bmbeta;\bmz_i)/\partial \bmbeta$ and  $\nabla^{2}\mL(\bmbeta;\bmz_i)$ $=\partial^2 \mL(\bmbeta;\bmz_i)/(\partial \bmbeta\partial \bmbeta^\top)$ are defined before equation (\ref{eq:subsample_loss_TS}) in Section \ref{sec:ensemble}. We further denote $\nabla^{\kappa}\mL(\bmbeta;\bmz_i)=\partial ^\kappa \mL(\bmbeta;\bmz_i)/(\underbrace{\partial \bmbeta^\top\otimes \cdots\otimes \partial \bmbeta^\top}_{\kappa})$ for any integer $\kappa\geq 3$, where $\otimes$ is  the Kronecker product.  In addition, let $\|\cdot\|_2$ be the vector $2$-norm or the matrix $2$-norm. Then we introduce the following technical conditions.


 (C1) Assume $\bmz_i\stackrel{iid}\sim \bmz$. In addition, assume that the true parameter vector $\bmbeta_0$ is an interior point of  a compact parameter space $\boldsymbol{\mathcal{B}}\subset\mathbb{R}^{p}$ and  $\bmbeta_0$ is the unique minimizer of $\rmE\{ \mL\lsk\bmbeta;\bmz\rsk\}$.

 (C2)  Assume that 
$\bmz\mapsto\mL(\bmbeta;\bmz)$ is measurable  given any $\bmbeta\in\boldsymbol{\mathcal{B}}$ and 
$\bmbeta\mapsto\mL(\bmbeta;\bmz)$ is four times continuously differentiable in $\boldsymbol{\mathcal{B}}$  for $P_{\bmz}$-almost every $\bmz$, where $P_{\bmz}=\rmP\circ {\bmz}^{-1}$ is the measure induced   by the random vector  $\bmz$.

 (C3)  Assume that  $\Sigma_{\bmbeta_0}=\Var\{\nabla\mL\lsk\bmbeta_0;\bmz\rsk\}$ and $V_{\bmbeta_0}=\rmE\{ \nabla^{2}\mL\lsk\bmbeta_0;\bmz\rsk\}$  are finite and positive definite.

 (C4) Assume that  $\rmE \sup_{\bmbeta\in \boldsymbol{\mathcal{B}}} \|\nabla^{\kappa}\mL(\bmbeta;\bmz)\|^2_2<\infty$ holds for $\kappa=1,2,3,4$.

 (C5) 
Assume that  $\rmE \|\nabla\mL\lsk\bmbeta_0;\bmz\rsk\|_2^4<\infty$ and
 $\rmE \|\nabla^2\mL\lsk\bmbeta_0;\bmz\rsk\|_2^4<\infty$ hold.

\noindent All the above conditions are mild and sensible, which are commonly used to establish the asymptotic properties and higher-order expansion of $M$-estimators (see, e.g.,   \citealp{rilstone1996second}, \citealp{van2000asymptotic},  \citealp{kim2016higher} and  \citealp{zou2021subbagging}).



\renewcommand{\theequation}{A.\arabic{equation}}
\setcounter{equation}{0}

\renewcommand{\thesubsection}{A.\arabic{subsection}}

\spacingset{1.12} 

\setlength{\bibsep}{4pt plus 0.3ex}
\bibliographystyle{apalike}
 \bibliography{reference}

\end{document}